\documentclass[aip,jmp,amsmath,amssymb,preprint,]{revtex4-1}
\usepackage[utf8]{inputenc}
\usepackage{amsmath}
\usepackage{amssymb}
\usepackage{graphicx}
\usepackage{subfigure}
\usepackage{color}

\begin{document}

\title{Rational evaluation of various epidemic models based on the COVID-19 data of China}

\author{Wuyue Yang$^{2\rm a)}$, Dongyan Zhang$^{3}$\footnote{Those authors contribute equally to this work.}, Liangrong Peng$^{4}$, Changjing Zhuge$^{3\rm b)}$, Liu Hong$^{1}$\footnote{Author to whom correspondence should be addressed. Electronic mail: hongliu@sysu.edu.cn (LH), zhuge@bjut.edu.cn (CZ)}\\
$^1$School of Mathematics, Sun Yat-sen University, Guangzhou, 510275, P.R.C.\\
$^2$Yau Mathematical Sciences Center, Tsinghua University, Beijing, 100084, P.R.C.\\
$^3$Beijing Institute for Scientific and Engineering Computing, Faculty of Sciences, Beijing University of Technology, Beijing 100124, P.R.C.\\
$^4$College of Mathematics and Data Science, Minjiang University, Fuzhou, 350108, P.R.C.}

\begin{abstract}
In this paper, based on the Akaike information criterion, root mean square error and robustness coefficient, a rational evaluation of various epidemic models/methods, including seven empirical functions, four statistical inference methods and five dynamical models, on their forecasting abilities is carried out. With respect to the outbreak data of COVID-19 epidemics in  China, we find that before the inflection point, all models fail to make a reliable prediction. The Logistic function consistently underestimates the final epidemic size, while the Gompertz's function makes an overestimation in all cases. Towards statistical inference methods, the methods of sequential Bayesian and time-dependent reproduction number are more accurate at the late stage of an epidemic. And the transition-like behavior of exponential growth method from underestimation to overestimation with respect to the inflection point might be useful for constructing a more reliable forecast. Compared to ODE-based SIR, SEIR and SEIR-AHQ models, the SEIR-QD and SEIR-PO models generally show a better performance on studying the COVID-19 epidemics, whose success we believe could be attributed to a proper trade-off between model complexity and fitting accuracy. Our findings not only are crucial for the forecast of COVID-19 epidemics, but also may apply to other infectious diseases.

\end{abstract}


\maketitle
\section{Background}
During the study of epidemics, one of the most significant and challenging problems is to forecast the future trends, like how many individuals might be infected each day, when the epidemics stop spreading, what kinds of policies and actions have to be taken and how they will influence the epidemics, and so forth \cite{Li2018,Lutz2019,Basu2013,Gingras2016}. The importance of epidemic forecast cannot be emphasized too much.

In the literature, various forecasting models/methods have been reported \cite{Chowell2016,Walters2018,Funk2019,Stocks2018,Roosa2019}. Among them, empirical functions, statistical inference methods and dynamical models (difference equations, differential equations, stochastic equations) are three major routines. Empirical functions, especially those with explicit forms, play an unreplaceable role in this field. They are simple, easily understandable, fast implemented and analyzable. The statistical inference methods are also highly welcomed, especially in the presence of a large amount of first-hand data. The basic goal of most statistical methods in epidemics is to estimate the basic/effective reproduction number, which serves as a key to evaluate the severe condition of an infectious disease. In dynamical models, the basic/effective reproduction number is transformed into reaction rate coefficients. Based on compartment assumptions on populations involved in epidemics, classical SI, SIR, SEIR model and many other generalized models are built. They show a great ability to correctly reproduce the basic features of an infectious disease, to uncover the hidden dynamics, like the numbers of exposed cases and asymptomatic carriers which are hard to be learnt from usual epidemiological investigation, to forecast the future trends of epidemics, as well as to evaluate the influence of diverse control policies and actions in quantity.

However, in the face of so many possible choices, which method is the best? Especially for the purpose of a reliable estimation on the epidemic trend in the future? In this paper, based on the COVID-19 data of Shanghai and other six provinces/cities in China during the spring of 2020, we explore this critical issue systematically. The performance of seven widely used empirical functions, four statistical inference methods and five dynamical models reported in the literature are compared in detail. The basic evaluation criteria, models/methods and data are summarized in Section II. Detailed analyses, comparisons and evaluations among various epidemic models are carried out in Section III. In Section IV, we forecast the first epidemic wave happening in Austria, Malaysia, Norway and Republic of Korea from March to June in 2020 as a further validation. The last section contains a conclusion and some brief discussions.

\section{Methods}
\subsection{Criteria and quantities for model evaluation}

It is far from a trivial problem to evaluate the forecast ability of various functions/models/methods in a rational way\cite{Tabataba2017,Funk2019,Roosa2019}. Many competing requirements should be considered at the same time. Here we employ three basic criteria as a general guiding principle, which can be measured through explicitly calculable quantities (see next section), \textit{i.e.}
\begin{itemize}
\item \textbf{Complexity v.s. Accuracy.} We seek for a well-balance between the model complexity and fitting accuracy. Neither too complicated models with numerous free parameters and unverified mechanisms, nor over-simplified models without sufficient capability to mimic the real situations is welcomed. This issue is also closely related to the over-fitting and under-fitting problems met in numerics.
\item \textbf{Fitting v.s. Prediction.} It is a very one-sided pursuit of the least fitting errors (measured by the mean square error, root mean square error, correlation coefficients, \textit{etc.}) for a predictive model, though it is often the case in most published works! In fact, there are tremendous evidences showing that the best fitting does not always lead to the best forecast (see Figs. S2 and S3 in SI for example). Just as an old Chinese proverb says, going too far is as bad as falling short. So we need to make a trade-off between the short-term best fittings and long-term promising predictions. A practical choice would be the statistical average of all possible results based on their weights (\textit{e.g.} the Boltzmann factor).
\item \textbf{Robustness v.s. Sensitivity.} On one hand, we hope our model is sensitive to parameter changes in order to model the influence of different situations and strategies, \textit{etc.} On the other hand, the model is expected to be robust (insensitive in other words) against perturbations arising from various sources, such as numerical errors, data noise, incomplete knowledge about epidemic mechanisms, \textit{etc.} Obviously, these two opposite pursuits can not be satisfied at the same time. Therefore, we turn to the reproducibility of key dynamical features (like the inflection point, half time) and the asymptotic stability (basic/effective reproduction number) of the model instead.
\end{itemize}

The three criteria above reflect the competition and compromise between model complexity and simplicity, over-fitting and under-fitting, short-term and long-term goals, robustness and sensitivity, as well as energetic and entropic, deterministic and statistical, local and global views. The over-emphasis of one aspect would lead to unsatisfactory predictions. As a perfect reflection of the central spirit of Confucianism -- Doctrine of the mean, which says that in all activities and thoughts one has to adhere to moderation, our three criteria provide a practical solution to overcome above difficulties both qualitatively and quantitatively. And thus they can be used for making a rational evaluation of different functions/methods/models for epidemic forecast and other related scientific problems.

To make model evaluation quantitatively, more concrete and easily measurable mathematical quantities are needed. Corresponding to each criterion discussed above, we consider the following factors:

(1) \textbf{The Akaike information criterion (AIC)} and its various modified versions, like AICc, AICu, QAIC, BIC, \textit{etc.} AIC was introduced by Japanese statistician Akaike in the early 1970s \cite{akaike1974new}. It is based on the concept of entropy, and incorporates the model complexity and its goodness of fit together.
\begin{equation}
AIC=2K-2\ln(L),
\end{equation}
in which $K$ is the total number of free parameters in a model, while $L$ is the likelihood function. Models with less free parameters and higher fitting accuracy will have lower AIC values. In this work, we use $AICc=-2\ln(L)/N+(N+K)/(N-K-2)$ proposed by  McQuarrie and Tsai \cite{Mcquarrie1998The} to remove the dependence on data size $N$. { According to Sugiura \cite{Sugiura1978Further}, when $K>N/40$, namely when the number of parameters is large in comparison to the number of time points, AICc should be adopted instead of AIC. The Akaike information criterion and its various modified versions have been widely used for model evaluation in the literature, e.g. see Refs. \cite{Martcheva2015An, Weston20why}. Especially, based on the early COVID-19 epidemic data of Wuhan, Weston \textit{et al.} made a preliminary comparison between the SIR and SEIR models by using the AIC value \cite{Weston20why}.}

(2) \textbf{The root mean square error (RMSE)}. The RMSE is extensively used to quantify the accuracy of regression models. It is defined as
\begin{equation}
RMSE=\sqrt{\frac{1}{N}\sum_{i=1}^N (x_i-y_i)^2},
\end{equation}
in which $N$ denotes the data size, $x_i$ and $y_i$ are the true values and predicted ones separately. {In this study, since we are dealing with the epidemic of a large province/city which includes at least millions of populations and hundreds of infected cases, the data (\textit{e.g.} the number of confirmed infected cases) follows a Gaussian distribution according to the central limit theorem, whose variance is generally proportional to the root square of the epidemic size. Therefore, RMSE is a natural quantity to characterize the prediction accuracy of various epidemic models.}

(3) \textbf{The robustness coefficient (RC)}. { There are plenty of ways to quantify model robustness. Here we adopt a simple definition based on the confidence interval. By randomly sampling the free parameter space, the best-fit values to the epidemic data and the $95\%$ confidence intervals are determined through Markov Chain Monte Carlo (MCMC) algorithms \cite{Chen2000}. Then the robustness coefficient is defined as the ratio between the smallest and the largest final epidemic size within $95\%$ confidence interval,
\begin{equation}
RC=\frac{the\,smallest\, final\, epidemic\, size\, among\, all\, predictions}{the\,largest\, final\, epidemic\, size\, among\, all\, predictions}.
\end{equation}
A RC value close to one indicates that the model predictions are consistent and reliable.

The robustness coefficient is closely related to the so-called ``nonidentifiability'' in the literature \cite{A2009Structural, Lintusaari2016On}, which means that a group of model parameters, giving the same good fit to the data but leading to completely different model predictions, cannot be uniquely determined during model calibration. The appearance of nonidentifiability may largely influence the reliability of model predictions and result in a relatively low robustness coefficient.

It should be noted in the current study AICc is calculated based on the training data, while the RMSE is calculated based on the test data, so that they are not synchronous. Furthermore, the robustness coefficient basically depends on the mathematical structure of the model, showing no direct correlation with AICc and RMSE.

\subsection{Model specification}
Far from complete, in the current study we collect seven empirical functions with explicit forms -- linear, exponential, logistic, Hill's, Gompertz's,  Richards', and generalized logistic functions \cite{zhao2019simple}; four statistical inference methods -- exponential growth, maximum likelihood, sequential Bayesian and time-dependent reproduction number \cite{obadia2012r0, li2020estimation}; as well as five dynamical models based on ordinary differential equations (ODEs) -- SIR, SEIR, SEIR-QD \cite{peng2020epidemic}, SEIR-AHQ \cite{Tang2020estimation} and SEIR-PO \cite{Zhang2020}  (see Fig. \ref{models}). Most of them have been frequently used in the literature to study the spreading of infectious diseases.

\begin{figure}[tp]
\[
\includegraphics[width=0.9\textwidth, height=1.2\textwidth]{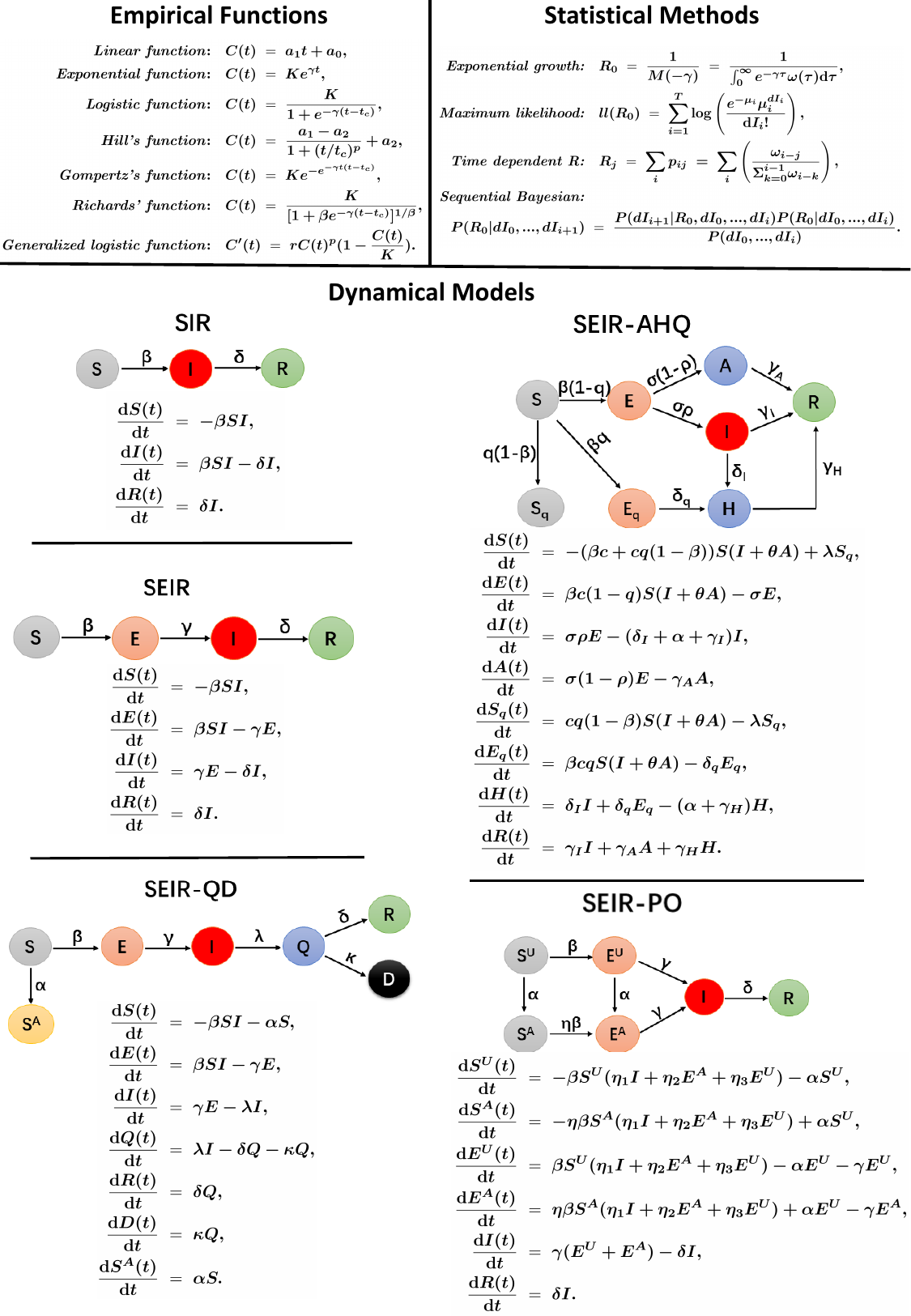}
\]
\caption{A summary on the empirical functions, statistical inference methods and dynamical models for epidemics evaluated in the current study. See Methods for details.}
\label{models}
\end{figure}

\subsubsection{Empirical functions}
To describe the growth of cumulative number of infected cases due to an infectious disease, like COVID-19, empirical functions in explicit forms are widely used \cite{zhao2019simple}. Here, the linear, logistic, exponential, Hill's, Gompertz's, Richards', and generalized logistic functions are summarized in the upper left corner of Fig.\ref{models}.

\subsubsection{Statistical methods}
Assuming a population to be totally susceptible, the basic reproduction number $R_0$ is defined as the average number of secondary infectious cases produced by one infectious case during a disease outbreak. The basic reproduction number $R_0$ plays a key role in studying the epidemics of infectious diseases. And many different statistical methods are designed for estimating $R_0$ \cite{li2020estimation},  some of which have been implemented with the ``$R_0$ package'' in R \cite{obadia2012r0}.

(1) \textbf{Exponential growth estimation}.

Exponential growth estimation method assumes that the number of infected cases increases exponentially, which is more suitable in the early stage of an epidemic. In this case, the basic reproduction number \cite{wallinga2007generation} is given by $$R_0=1/M(-\gamma)={1}/{\int_0^{\infty}e^{-\gamma\tau}\omega(\tau)d\tau},$$ 
where $\gamma$ is the growth rate and $M$ is the moment generating function of the generation time distribution $\omega(\tau)$. The latter is generally assumed to follow the Gamma distribution.

(2) \textbf{Maximum likelihood estimation}.

This method assumes the number of infected cases generated from the first case follows the Poisson distribution, whose mean is directly proportional to the basic reproduction number and can be estimated by using the maximum likelihood method \cite{forsberg2008likelihood},
$$
ll\left( R_{0} \right) = \sum\limits_{i = 1}^T {\log \left( {\frac{{{e^{ - {\mu _i}}}{\mu _i}^{{dI_i}}}}{{{dI_i}!}}} \right)},~
{\mu _i} = R_0\sum\limits_{k = 1}^i {{dI_{i - k}}{\omega_k}}.
$$
Here $ll(R_0)$ is the likelihood depending on $R_0$. $\mu_i$ and $dI_i=I_i-I_{i-1}$ are the number of daily new infected cases and incident cases at discrete time point $i$, $w_i$ is the generation time distribution. This method also requires the period during which the exponential growth is happening to be identified from the data by statistical tools.

(3) \textbf{Sequential Bayesian method}.

The sequential Bayesian method, or real-time Bayesian, starts with a non-informative prior and tries to predict the posterior distribution of the basic reproduction number $R_0$ by referring to the Bayesian formula \cite{bettencourt2008real},
$$
P\left( R_0| dI_0, \cdots, dI_{i+1} \right) = \frac{P\left( dI_{i+1}| R_0, dI_0, \cdots, dI_{i} \right) P\left( R_0| dI_0, \cdots, dI_{i} \right) }{P\left(dI_0, \cdots, dI_{i} \right) },
$$
where $P\left( dI_{i+1}| R_0, dI_0, \cdots, dI_{i} \right)$ is the likelihood of observing incident cases at time $i+1$ given the value of $R_0$ and past observations of incident cases from time $0$ to $i$, $P\left( R_0| dI_0, \cdots, dI_{i} \right)$ is a prior distribution of the basic reproduction number, and $P\left(dI_0, \cdots, dI_{i} \right)$ is the joint probability of observing the incident cases. The number of daily new infected cases is also assumed to be Poisson distributed with the mean $\mu_i=dI_{i-1}e^{\gamma(R_0-1)}.$

(4) \textbf{Estimation of time dependent reproduction numbers}.

This method computes the basic reproduction numbers by averaging over all transmission networks compatible with observations\cite{wallinga2004different}. The relative likelihood ${p_{ij}}$, that a case onset at time $i$ was infected by a case onset at time $j$, is given by ${p_{ij}}={\omega_{i-j}}/{\sum_{k=0}^{i-1}\omega_{i-k}}$.
Consequently, the time-dependent effective reproduction number for case $j$ is defined as ${R_j} = \sum_i {p_{ij}}$, and the basic reproduction number is the average of all $R_j$, i.e. ${R_0} =\frac{1}{T}\sum_{j=1}^TR_j$.

\subsubsection{ODE-based dynamical equations}

Without considering time delay and spatial heterogeneity, ordinary differential equations are the most widely used models for describing the spreading process of epidemics. Here we summarize five different dynamical models reported in the literature for studying COVID-19.

(1) \textbf{SIR model}

The classical SIR model divides populations into three compartments, that is {\em susceptible}, {\em infectious} (with infectious capacity and not yet recovered) and {\em recovered} cases (recovered and not be either infectious or infected once again) denoted by $S(t)$, $I(t)$ and $R(t)$ separately. As shown in Fig.\ref{models}, the coefficients $\beta$ and $\delta$ represent the infection rate and recovery rate separately.

(2) \textbf{SEIR model}

To account for the infected cases which are still in a latent period and not yet being infectious, a new {\em exposed} population $E(t)$ is introduced in SEIR model \cite{wang2020phase}, in which a new coefficient $\gamma$ denotes the transition rate from exposed individuals to the infected.

(3) \textbf{SEIR-QD model}

To take the effects of quarantine and self-protection into consideration, Peng \textit{et al.} \cite{peng2020epidemic} proposed to generalize the classical SEIR model by introducing a new {\em quarantined} state between {\em infectious} and {\em recovery}. The numbers of {\em death} and {\em unsusceptible} are denoted as $D(t)$ and $S^A(t)$ separately.
In SEIR-QD model, the coefficients $\alpha, \lambda, \delta, \kappa$ denote the protection rate of susceptible individuals, the transition rate from infectious individuals to the quarantined infected class, the recovery rate and death rate, respectively.

(4) \textbf{SEIR-AHQ model}

To incorporate appropriate compartments relevant to interventions such as quarantine, isolation and treatment, Tang \textit{et al.} \cite{Tang2020estimation} generalized the SEIR model. They stratified the populations as {\em susceptible} ($S$), {\em exposed} ($E$), infectious but not yet symptomatic
({\em pre-symptomatic}) ($A$), {\em infectious} with symptoms ($I$), {\em hospitalized} ($H$) and {\em recovered} ($R$)
compartments, and further included {\em quarantined susceptible} ($S_q$) and {\em isolated
exposed} ($E_q$) compartments.

In SEIR-AHQ model, the parameters $\{c,\beta,q,\sigma,\lambda,\rho,\delta_I,\delta_q,\gamma_I,\gamma_A,\gamma_H,\alpha\}$ represent the contact rate, probability of transmission per contact, quarantined rate of exposed individuals, transition rate from exposed individuals to the infected, release rate of uninfected contacts from quarantine, probability of having symptoms among infected individuals, transition rate of symptomatic infected individuals to the quarantined infected class, transition rate from quarantined exposed individuals to the quarantined infected class, recovery rates of symptomatic infected individuals, asymptomatic infected individuals and quarantined infected individuals, as well as disease-induced death rate.

(5) \textbf{SEIR-PO model}

By incorporating the public opinion on COVID-19, Zhang \textit{et al.} \cite{Zhang2020} further classified the populations of {\em susceptible} and {\em exposed} in SEIR model into \textit{unconscious} ($S^U, E^U$) and \textit{conscious} ($S^A, E^A$) based on their different knowledge on epidemics and self-protection.

In SEIR-PO model, the parameters $\{\gamma,\delta,\beta, \eta,\eta_1,\eta_2,\eta_3,\alpha\}$ denote the transition rate from exposed individuals to the infected, recovery rate of infected individuals, infection rate of unconscious susceptible population, reduced infection ratio of conscious susceptible individuals, effective infection factors of infectious individuals, unconscious and conscious exposed individuals, as well as the spreading rate of knowledge about COVID-19 among individuals.

\subsection{Data}

To make a quantitative comparison, here we focus on the outbreaks of COVID-19 caused by the novel coronavirus -- SARS-CoV-2, which currently spreads severely worldwide. We download the data of daily reported confirmed infected cases $C(t)$ from the China CDC (http://www.chinacdc.cn/). As a first example, the public epidemic data of Shanghai is studied. Shanghai is in the east of China and is considered as one of the best controlled cities in China during the battle against COVID-19. A forty-day period from Jan. 20th, 2020 to Feb. 28th, 2020 is equally divided into four sequential series by every ten days.


Similarly, a larger data set is collected during almost the same time period in six different regions in China selected mainly according to their geographic locations (see Fig. \ref{china-map}), including Heilongjiang province (northeast China), Tianjin (northern), Guangdong province (southern), Chongqing (southwest), Hunan province (central) and Xiaogan city in Hubei province (central, the city with the second largest reported infected populations). Each data set is divided into two instead of four for simplicity.

\begin{figure}[ht]
\[
\includegraphics[scale=0.6]{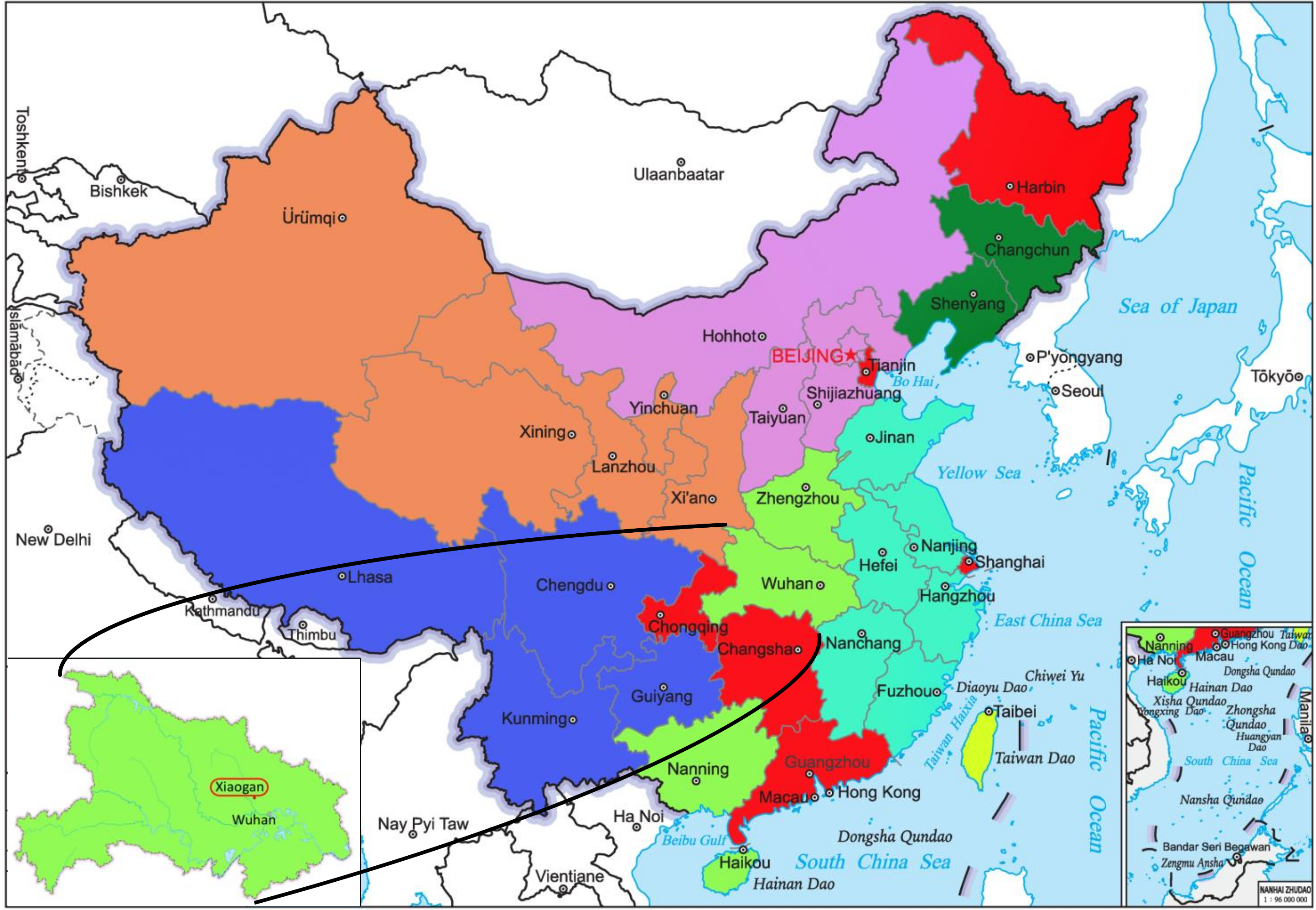}
\]
\caption{The geographic locations of seven provinces/cities studied in the current research (colored in red). The maps are depicted based on the standard maps (GS(2019)1647,GS(2019)3333) from Standard Map Service (bzdt.ch.mnr.gov.cn) by Ministry of Natural Resources of the People’s Republic of China. }
\label{china-map}
\end{figure}

\subsection{Parameter estimation}

The fitting procedure of all empirical functions is done through the nonlinear fitting mode in Origin 2019.

As to four statistical methods, since we aim at making predictions on the progression of epidemics, we need to combine them with further assumptions on the dynamics. A widely adopted one is the exponential growth, which assumes the number of infected populations grows exponentially with the time and the exponent $\gamma$ is correlated with the basic reproduction number as $R_0=1/{\int_0^{\infty}e^{-\gamma\tau}\omega(\tau)d\tau}$.
In the current study, we assume the generating time distribution $\omega(t)$ obeys the Gamma distribution $\Gamma(k,\theta)$ (see Fig. S1 in SI), whose moment generating function is explicitly known as
$M(t) = {\left( {1 - t\theta } \right)^{ - k}}, \forall t < 1/\theta$.
From it, we immediately see $\gamma=(R_0^{1/k}-1)/\theta$. Then inserting $\gamma$ into either the recurrence formula $\mu_i=dI_{i-1}e^{\gamma(R_0-1)}$ (no free parameter) or the Logistic function (two more free parameters), the progression of epidemics is fitted and predicted.

The unknown parameters involved in dynamical models are estimated respectively by fitting the models to the epidemic data through either the standard nonlinear least-squares approach \cite{peng2020epidemic} or the Markov-Chain Monte-Carlo (MCMC) algorithms. The MCMC algorithms are widely used in this field to sample the parameter space and to fit the model to the data \cite{Chen2000, Tang2020estimation}. The MCMC is performed through an adaptive Metropolis-Hastings algorithm, which is implemented in the R package POMP \cite{King2016Statistical}. 80,000 iterations with a burn-in of the first 50,000 iterations are carried out, where non-informative uniform distributions are chosen as the prior distributions. From the posterior distributions, we obtain the best-fit values and their $95\%$ confidence intervals.

\section{Results}
In this part, we apply our three basic criteria to evaluate several widely used models/methods in the field of epidemics. By fitting  models to the training data set of Shanghai with varied time spans, their forecasting abilities are compared and quantified through AICc, RMSE and RC values (see Fig. \ref{shanghai} and Table \ref{table1}). Analogously, the results for other six cities/provices are summarized in Fig. \ref{six-cities} and Table \ref{table2}.

Based on extensive numerical explorations of models for COVID-19 with epidemic data of China, our main findings are summarized as follows:

(1) \textbf{\em The model with the least RMSE can be picked out based on AIC and RC.} An astonishing finding of our current study is that the model with the least RMSE to the test data can be easily picked out by examining the values of AICc and RC, while the latter two depend on the model and training data only! As we learned in the previous section, the lower the AICc value is, the better trade-off between model complexity and fitting accuracy is achieved. Meanwhile, a medium RC value ($0.5-0.9$ in general) could take the requirements on model robustness and sensitivity into consideration at the same time. Among 27 groups of epidemic data (9 cases times 3 groups of models) under comparison, the AICc value helps to find out 18 models with the least RMSE to the test data, and the rest 9 models all have the second lowest AICc values (see Table. \ref{table1} and Table. \ref{table2} for details). This finding is consistent with previous reports based on Ebola epidemics that reactive models are of better performance for short-term weekly incidence if they have few parameters. \cite{2018Viboud}

(2) \textbf{\em Sigmoid functions are more suitable for epidemic forecast.}
Linear and exponential functions are not suitable for describing epidemic data in general, while Hill's, Logistic, generalized Logistic, Gompertz's and Richards' functions can well capture the typical S-shaped curve for the cumulative infected cases.

(3) \textbf{\em At the early stage of an epidemic, no model can make long-term reliable forecast.} With respect to very limited data in the early stage of an epidemic, there is no way to tell which model is superior than the rest. They may either overestimate or underestimate the epidemic size in an unpredictable way. Since the model with fewer parameters is more robust, we suggest adopting either the exponential function or even the linear function, though their valid regions are quite narrow. A more elegant way is to combine the knowledge of the basic reproduction number derived from statistical methods and the forecast ability of exponential or logistic functions. However, it should be noted that during the early stage the variance of the derived basic/effective reproduction number is generally very large (see Fig. S1 in SI), making long-term reliable forecast almost impossible.

(4)\textbf{\em The inflection point is crucial for forecast.} The inflection point plays an essential role in forecast. It was suggested by Zhao \textit{et al.} \cite{zhao2019simple} in Zika research that, when the epidemic passes the inflection point, predictions on the final epidemic size by the sigmoid empirical functions, such as Logistic, Gompertz's and Richards' functions, will converge to the true values. Here we basically reproduce their results. As shown in the last row of Fig. \ref{shanghai}, the RMSE of predictions on the COVID-19 epidemic data of Shanghai decays in an exponential way with respect to the size of training data, meanwhile the RMSE of fitting keeps nearly unchanged. Interestingly, before Jan. 31st which is also the inflection point of Shanghai, all functions fail to make a reliable prediction (and some functions fail even earlier) and their RC values drop to zero rapidly.

\begin{figure}[htbp]
\centering
{
\begin{minipage}[t]{1\linewidth}
\centering
\includegraphics[width=0.95\textwidth, height=0.7\textwidth]{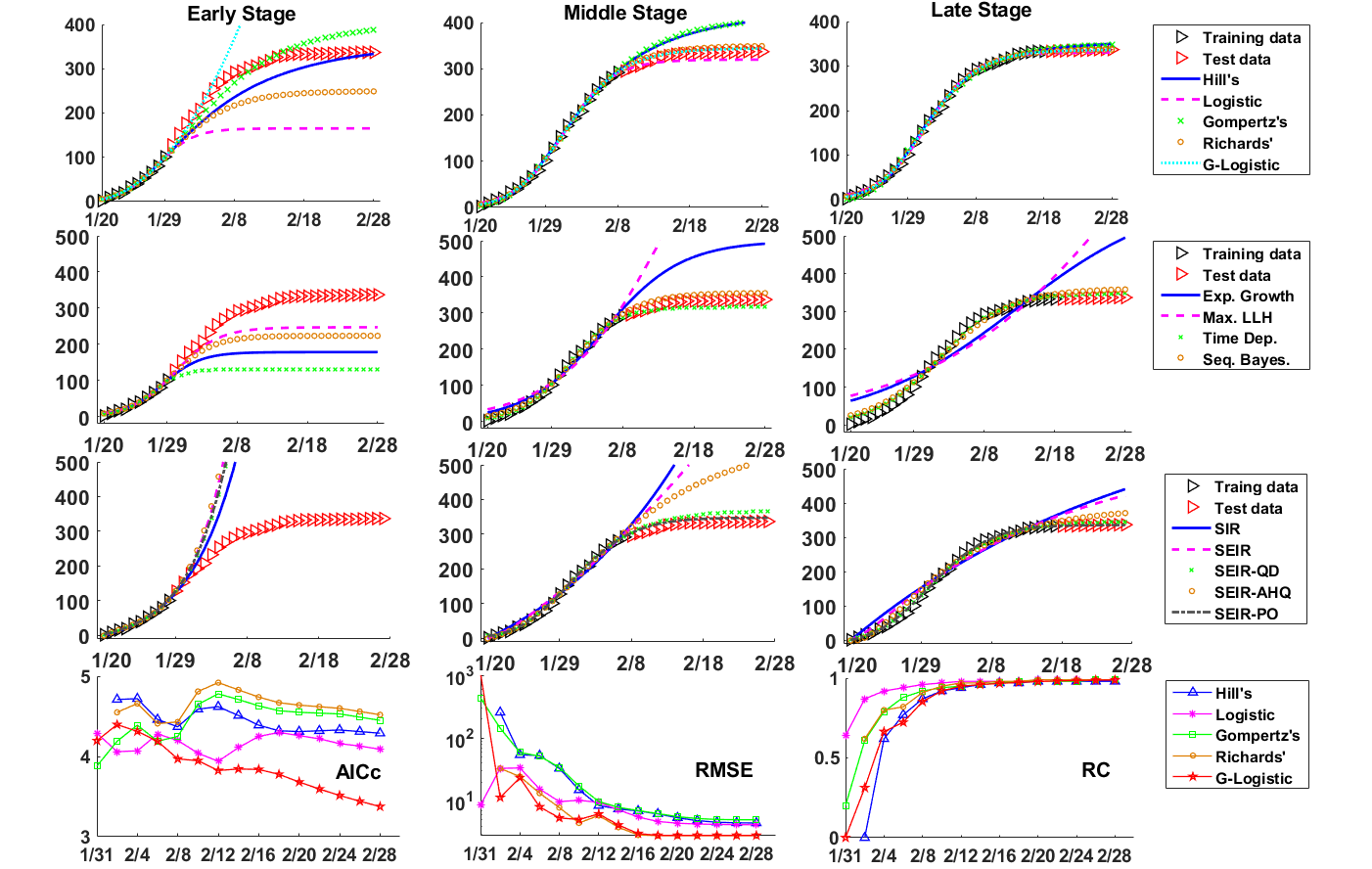}
\end{minipage}%
}%
\caption{Forecast of the COVID-19 epidemic in Shanghai from 01/20/2020 to 02/28/2020 based on the data of first 10 (early), 20 (middle) and 30 (late) days respectively. The first three panels give the results of (upper) five explicit functions, (middle) four different statistical inference methods combined with the Logistic function (the exponent $\gamma$ derived from $R_0$), (lower) and five ODE models. The one with the smallest RMSE to the training data is drawn. The last row shows the variations of AICc (for training data), RMSE (for test data) and RC for four explicit functions with respect to different sizes of training data set (from Jan. 20th to the date as marked).}
\label{shanghai}
\end{figure}

\begin{table}[ht]
\[
\includegraphics[scale=0.8]{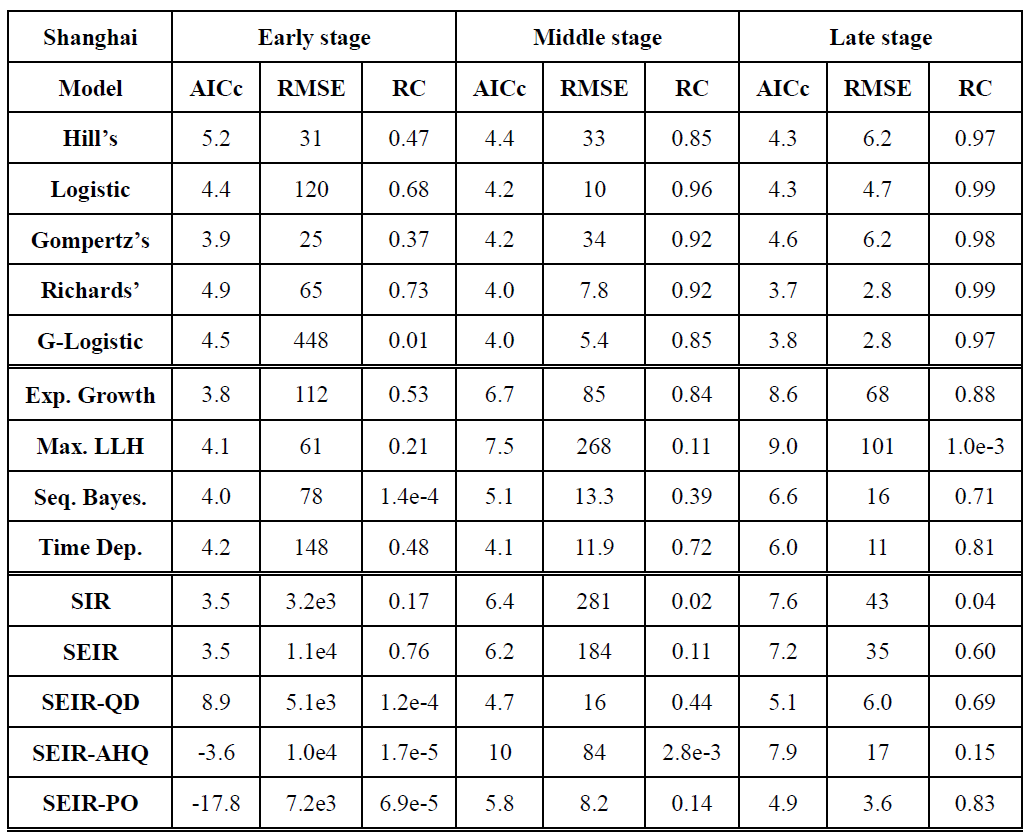}
\]
\caption{Summary of AICc (for training data), RMSE (for test data) and RC values for different models calculated based on the epidemic data of Shanghai. Note the negative AICc values result from the fact that data points are fewer than the free model parameters.}
\label{table1}
\end{table}

(5) \textbf{\em The Logistic function underestimates the epidemic size while Gompertz's function overestimates it.} In all nine cases (including three cases for Shanghai), we notice the Logistic functions always underestimate the total number of infected cases, while the Gompertz's function makes an overestimate (see Fig. \ref{six-cities}A). This finding needs further exploring and would be useful for estimating the lower and upper bounds for the real total infected populations, though it still requires further validation. The results of the other three functions are not so consistent and their goodness-of-fit varies from case to case.

(6) \textbf{\em Methods of sequential Bayesian and time-dependent reproduction number are more accurate at the late stage of an epidemic.} For statistical methods, since sequential Bayesian and time-dependent reproduction number methods take the non-constant nature of the effective reproduction number with the progression of epidemics into consideration (see Fig. S1 in SI), their predictions appear to be more accurate than the exponential growth and maximum likelihood methods in the late stage (see Fig. \ref{six-cities}B). In addition, the sequential Bayesian method seems to be less robust than the time-dependent reproduction number method. The latter inherits the merit of Logistic function by slightly underestimating the true epidemic size. The nice performance of Bayesian method has been observed for Ebola too, in which ensemble Bayesian method outperformed other 8 methods including Logistic function and SEIR model \cite{2018Viboud}. It is further observed that the basic reproduction number $R_0$ estimated by the exponential growth method exhibits a transition from overestimation to underestimation with respect to the inflection point, which is in accordance with the S-shaped curve for the total infected populations. As a consequence, with respect to the early stage data of Shanghai COVID-19 epidemic, the exponential growth method combining with the Logistic function makes an underestimation on the final epidemic size, and a contrary overestimation based on accumulated data in the late stage. Finally, we find the maximum likelihood method overestimates the epidemic size to a large extent in all seven cases, indicating this method may not be suitable for studying COVID-19 epidemics.

(7) \textbf{\em The SEIR-QD and SEIR-PO models are suitable for modeling COVID-19 epidemics.} The dynamical models generally require more training data to achieve a reliable forecast than empirical functions, since the former usually involves more free parameters and more complicated mathematical structure. Based on their performance, the dynamical models can be classified into three groups. As shown in Fig. \ref{shanghai} and Fig. \ref{six-cities}C, The classical SIR model and SEIR model seem to be inadequate to describe the outbreak of COVID-19, especially the final equilibration phase. Contrarily, the SEIR-AHQ model involves too many free parameters as reflected through the large AICc value. As a consequence, its robustness is also the poorest among all five models. The SEIR-QD and SEIR-PO models are two suitable ones for modeling COVID-19 by appropriately incorporating the effects of quarantine and self-protection.

It is noted that all of our above statements have to be remained specific to COVID-19 in the countries where data were fitted. Application to other scenarios should be done with great care and requires further exploration.

\begin{figure}[htbp]
\centering
\begin{minipage}[t]{1\linewidth}
\centering
\includegraphics[width=0.7\textwidth, height=0.35\textwidth]{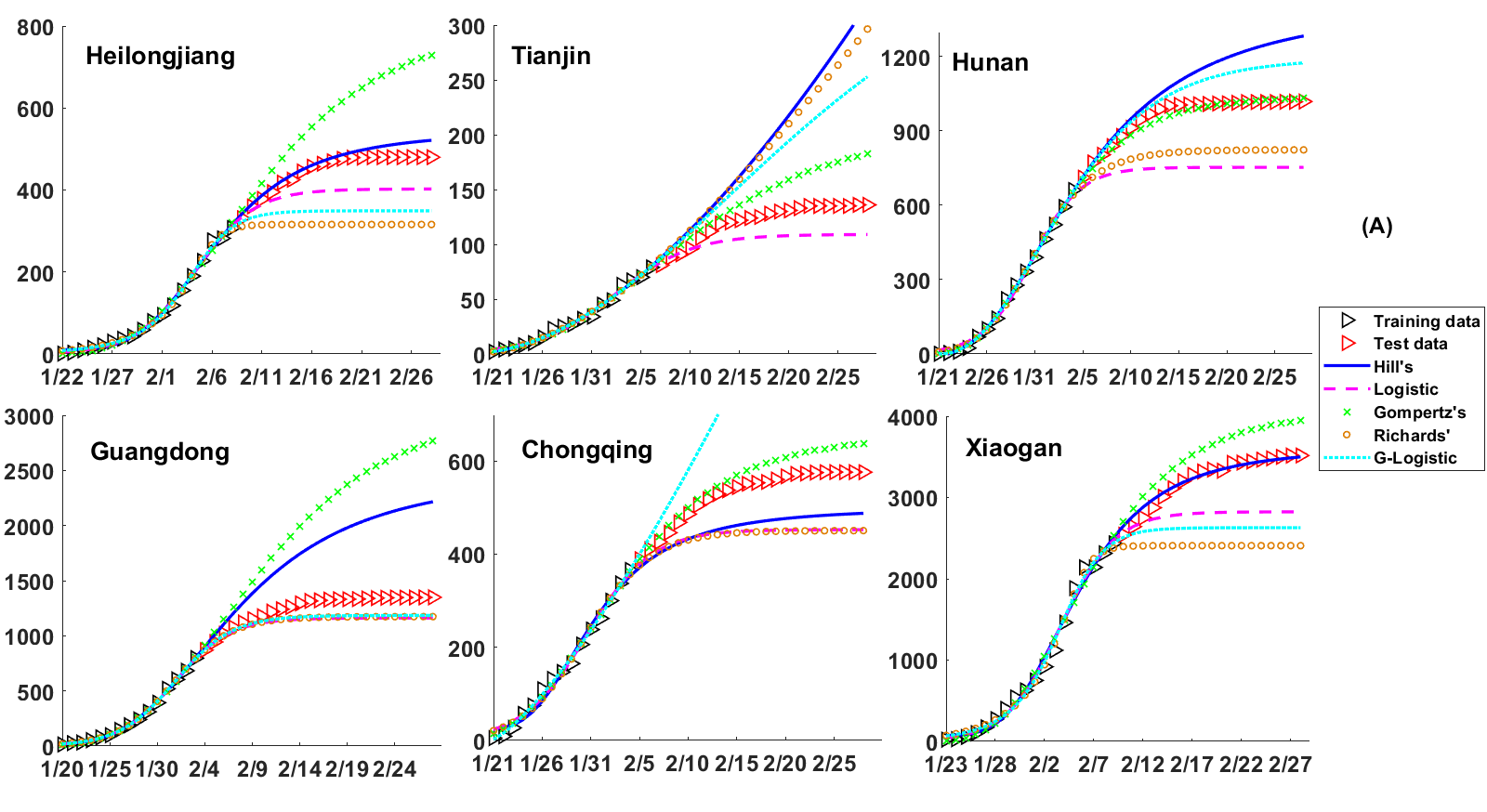}
\end{minipage}%

\begin{minipage}[t]{1\linewidth}
\centering
\includegraphics[width=0.7\textwidth, height=0.35\textwidth]{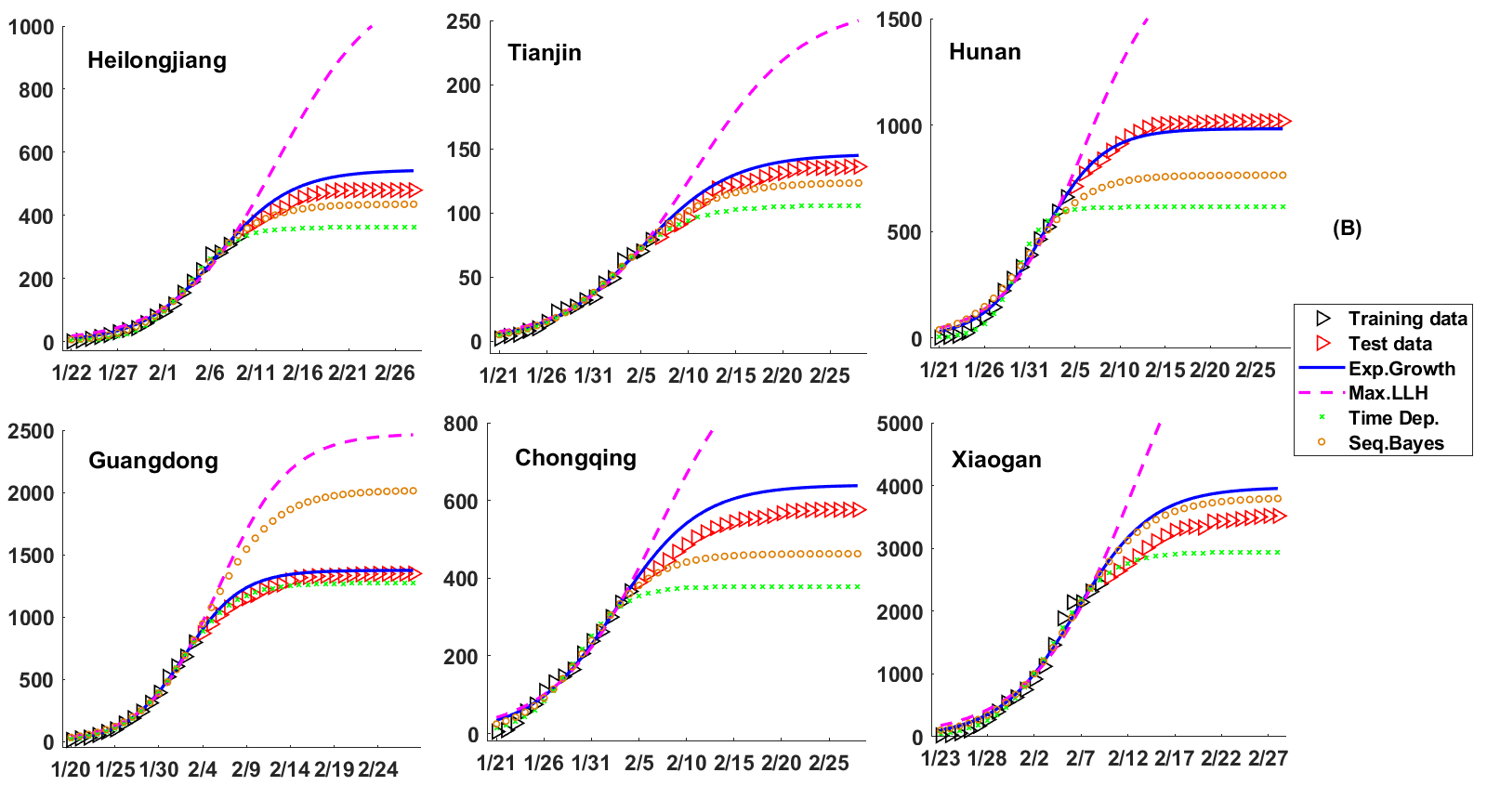}
\end{minipage}%

\begin{minipage}[t]{1\linewidth}
\centering
\includegraphics[width=0.7\textwidth, height=0.35\textwidth]{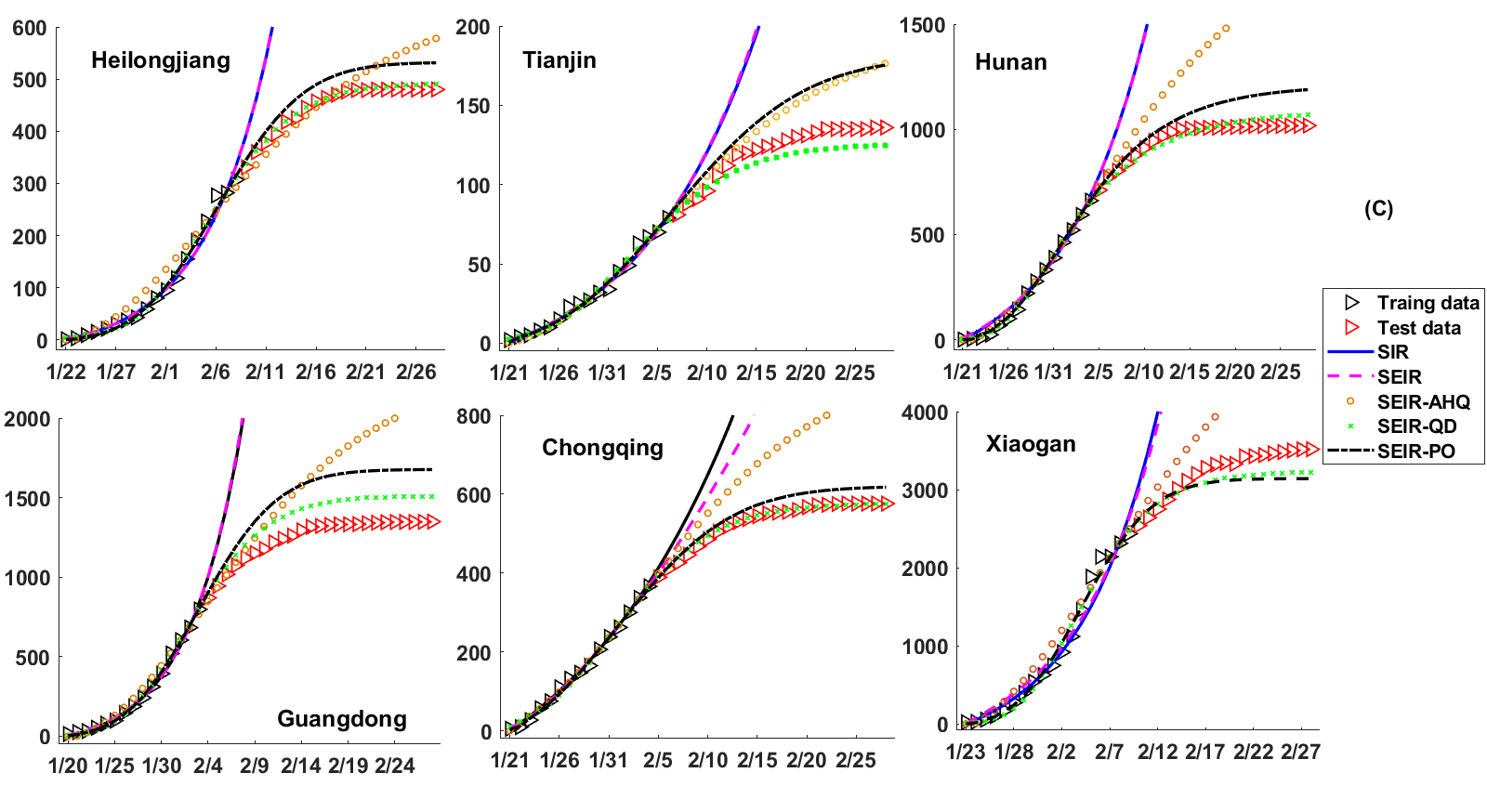}
\end{minipage}%
\caption{Forecast of COVID-19 epidemics in Heilongjiang province (data for first 18 days of 38 in total are used for modeling fitting (training set), while the rest 20 data points are used for validation (test set)), Tianjin (17/39), Hunan province (15/39), Guangdong province (15/40), Chongqing (15/39) and Xiaogan city (18/37) in Hubei province (central, city with the second largest reported infected populations). The upper panel (A) shows the results of five empirical functions, the middle one (B) for four different statistical inference methods combined with the Logistic function (the exponent $\gamma$ derived from $R_0$), while the lower panel (C) gives the results of five ODE models. The one with the smallest RMSE to the training data is drawn.}
\label{six-cities}
\end{figure}

\begin{table}[ht]
\[
\rotatebox{-90}{\includegraphics[scale=0.8]{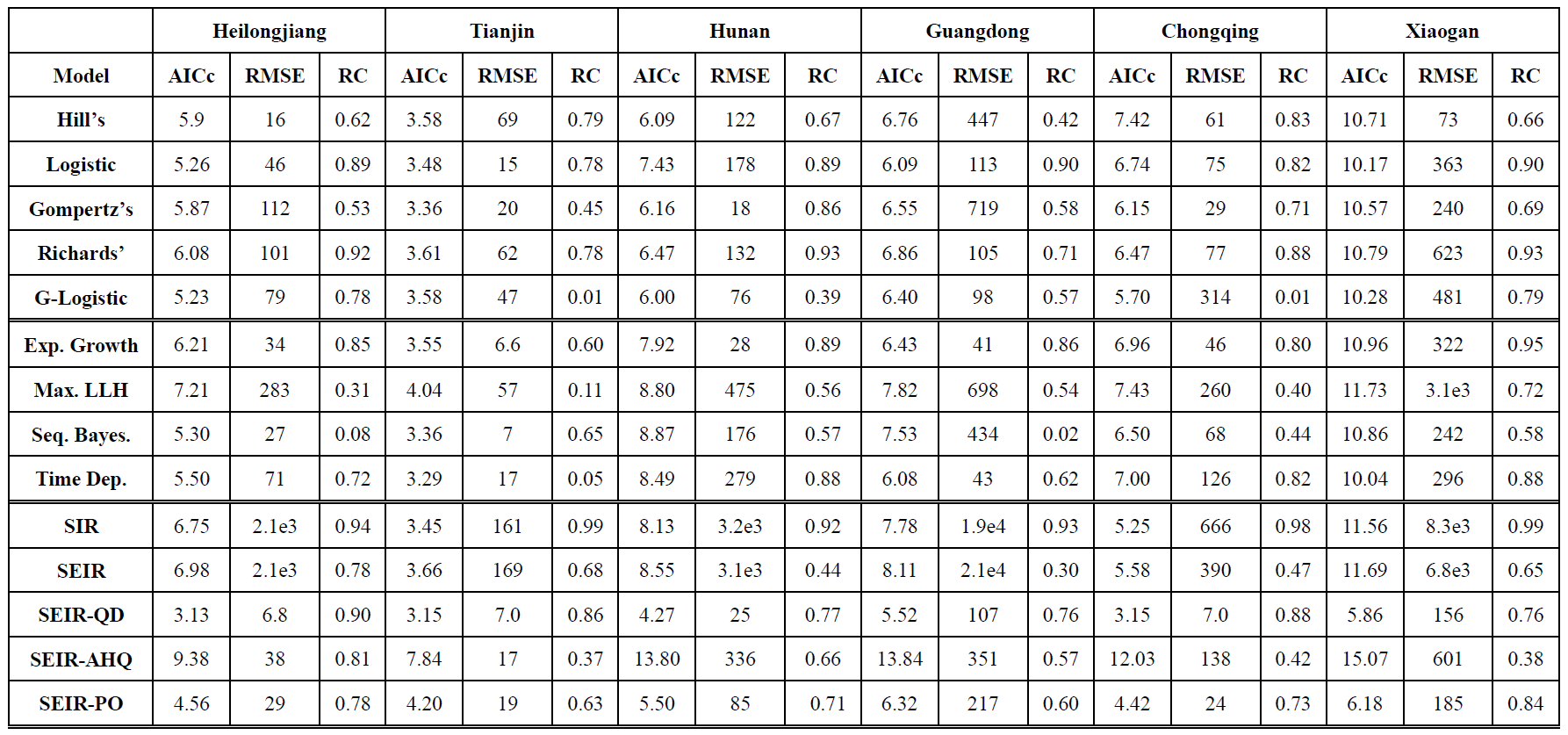}}
\]
\caption{Summary of AICc (for training data), RMSE (for test data) and RC values for different models calculated based on the epidemic data of six provinces/cities in China.}
\label{table2}
\end{table}

\section{Epidemic trends in other countries}
Based on our previous evaluations on different epidemic models and methods, we attempt to predict the epidemic trends in several other countries other than China, which also serves as a further validation of our statements.

In Fig. \ref{forecast}, we look into four examples -- the first epidemic wave of Austria, Malaysia, Norway and Republic of Korea, which have been randomly selected from their representative categories (four groups of countries clustered by the k-means algorithm according to their diverse control policies, unpublished data). In each case, the effective reproduction number $R_t$ is derived from the method of time-dependent reproduction number. It is well-known that the fact $R_t>1$ indicates a rapid spreading of the coronavirus. In particular, an astonishing peak of $R_t$ is observed for Republic of Korea on Feb. 19th, which might be attributed to the appearance of super-spreaders.  With respect to the total confirmed infected cases reported in these four countries, data after the inflection point are predicted by the Gompertz's and Logistic functions separately, since according to our previous findings the Gompertz's function usually overestimates the final epidemic size while the Logistic function underestimates it. Here only one exception is observed (the test data of Malaysia is larger than the prediction by the Gompertz's function), which is likely caused by the early occurrence of the second epidemic wave. As a conclusion, we can still believe the Gompertz's function and the Logistic function provide reasonable upper and lower bounds for the total confirmed infected case at least for the near future.

\begin{figure}[ht]
\[
\includegraphics[scale=0.5]{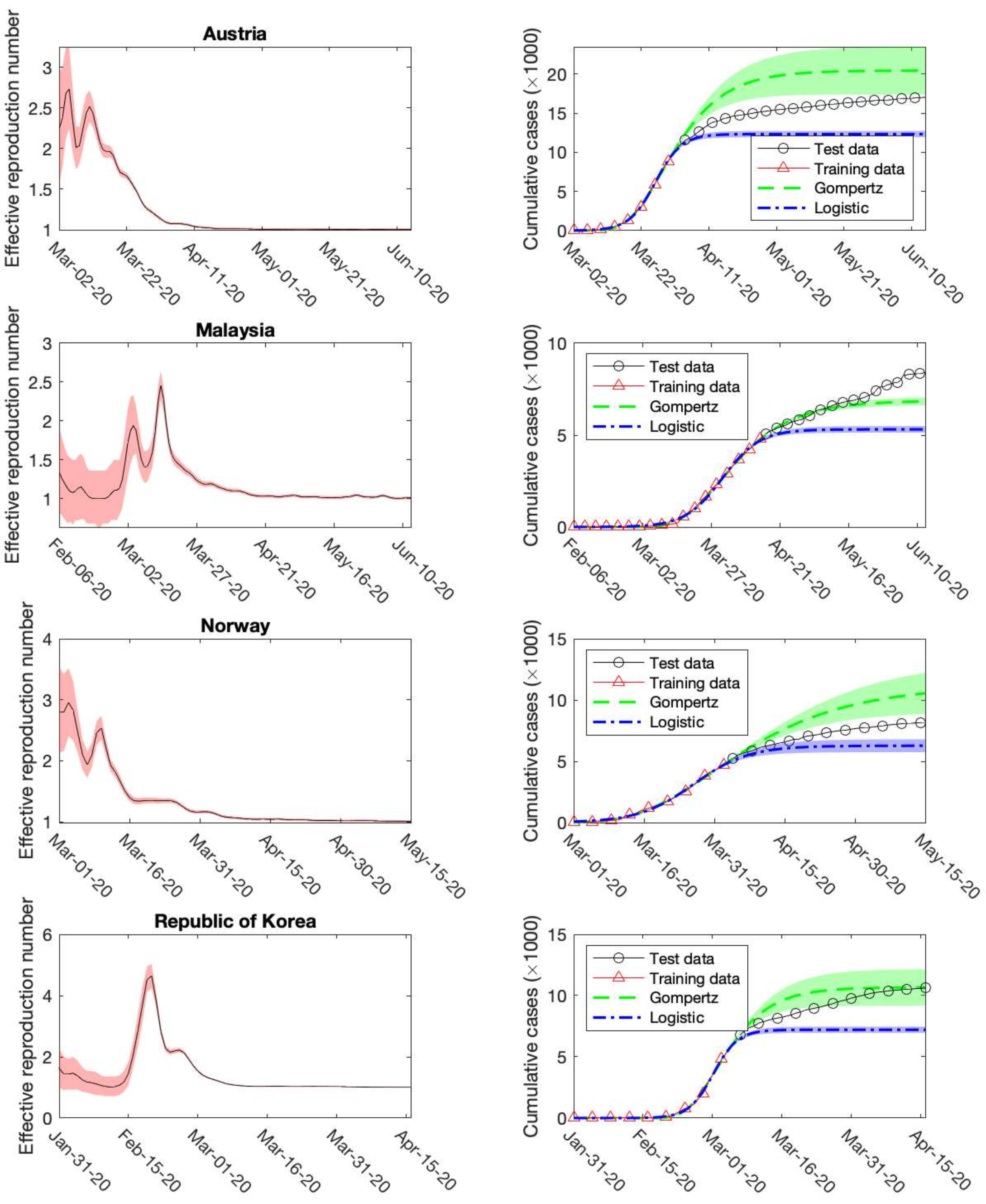}
\]
\caption{Forecast of the COVID-19 epidemics in Austria, Malaysia, Norway and Republic of Korea in 2020. Left column shows the effective reproduction number derived from the method of time-dependent reproduction number, while the right column gives the reported confirmed cases and predicted ones by using Gompertz's and Logistic functions.}
\label{forecast}
\end{figure}

\section{Conclusion and Discussion}

In this paper, based on the COVID-19 data of seven provinces/cities in China during the spring of 2020, we make a systematical investigation on the forecast ability of seven widely used empirical functions, four statistical inference methods and five dynamical models reported in the literature. We highlight the significance of a good balance between model complexity and accuracy, over-fitting and under-fitting, as well as model robustness and sensitivity for model performance. Quantitative analyses are made with respect to the Akaike information criterion, root mean square error and robustness coefficient.

Through extensive simulations and detailed comparisons, we find that the inflection point plays a crucial role for making reliable forecasts, in agreement with previous reports \cite{zhao2019simple}. The RMSE of model prediction decays exponentially with respect to the size of training data set, while the model robustness characterized through the variance of final epidemic size also approaches to unity rapidly after the inflection point. Furthermore, the forecast abilities of several epidemic models are also closely related to the inflection point. For example, the estimated basic reproduction number $R_0$ by the exponential growth method exhibits a transition from overestimation to underestimation with the increase of the training data set, and the inflection point acts as the demarcation.

We notice the Logistic functions always underestimate the total number of infected cases, while the Gompertz's function makes an overestimation in all cases we studied. Generalized Logistic, Hill's and Richards' functions do not have such a consistency. Since the sequential Bayesian and time-dependent reproduction number methods take the non-constant nature of the effective reproduction number with the progression of epidemics into consideration, we think they are more accurate than the exponential growth and maximum likelihood methods especially in the late stage of an epidemic. The transition of exponential growth method from underestimation to overestimation with respect to the inflection point could be quite useful for constructing a more reliable forecast. Towards the dynamic models based on ODEs, it is observed that the SEIR-QD and SEIR-PO models generally show a better performance than the other three, highlighting the significance of a trade-off between model complexity and fitting accuracy. The success of the former two models could also be attributed to the inclusion of self-protection and quarantine during the progression of COVID-19 epidemics.

There are many factors, like changing the reporting rate, increasing the testing capacity, improving the social awareness and self-protection, promoting vaccine injection and herd immunity, may affect the epidemic to a great degree. Generally, these factors are highly time- and policy-dependent, varied from region to region, may or may not be fully considered in various models. In the current study, the influence of these factors has not been thoroughly examined, and we call the readers' attention to this point. Furthermore, besides ODE models, partial differential equations \cite{Martcheva2015An}, stochastic equations \cite{Ma2009Dynamical} and time-delayed equations \cite{Yue2020modeling} have been applied to this field too. For example, it has been claimed that ``stochastic models should be preferred to deterministic models in most circumstances because they afford improved accounting for real variability and increased opportunity for quantifying uncertainty'' \cite{King2014Avoidable}. How to generalize our current research to these models would be of great value. Interested readers may refer to \cite{Konishi2008Information, Stocks2018Model, Gibson2018Comparison} for further details.

\subsection*{Authors' contributions}
LH and CZ designed the project. WYY, DYZ collected the data. All
authors analyzed the data. LH, CZ, WYY, LRP wrote the manuscript, and all authors reviewed it.

\subsection*{Declaration of Competing Interest}
The authors declared no competing interests.

\subsection*{Acknowledgements}
The authors acknowledged the financial supports from the National Natural Science Foundation of China (Grants No. 21877070, 11801020), Startup Research Funding of Minjiang
University (mjy19033), the Natural Science Foundation of Fujian Province of China (2020J05172), and Special Pre-research Project of Beijing University
of Technology for Fighting the Outbreak of Epidemics. Zhuge would like to thank
Dr. Yi Wei for his support on data collection.

\section*{Appendix A. Supplementary data}
Supplementary material related to this article can be found, in the
online version, at doi: XXX

\bibliographystyle{unsrt}
\bibliography{bibfile}

\end{document}